\documentstyle[twoside]{article}
\catcode`\@=11
\long\def\@makefntext#1{
\protect\noindent \hbox to 3.2pt {\hskip-.9pt
$^{{\eightrm\@thefnmark}}$\hfil}#1\hfill}       

\def\@makefnmark{\hbox to 0pt{$^{\@thefnmark}$\hss}}    

\def\ps@myheadings{\let\@mkboth\@gobbletwo
\def\@oddhead{\hbox{}
\rightmark\hfil\eightrm\thepage}
\def\@oddfoot{}\def\@evenhead{\eightrm\thepage\hfil
\leftmark\hbox{}}\def\@evenfoot{}
\def\sectionmark##1{}\def\subsectionmark##1{}}

\oddsidemargin=\evensidemargin
\newcounter{sectionc}\newcounter{subsectionc}\newcounter{subsubsectionc}
\renewcommand{\section}[1] {\vspace{12pt}\refstepcounter{sectionc}
    \noindent
    {\tenbf\thesectionc. #1}\par\vspace{5pt}}
\renewcommand{\subsection}[1] {\vspace{12pt}\refstepcounter{subsectionc}
    \noindent
    {\bf\thesectionc.\thesubsectionc. {\kern1pt \bfit #1}}\par\vspace{5pt}}
\renewcommand{\subsubsection}[1] {\vspace{12pt}\refstepcounter{subsubsectionc}
    \noindent{\tenrm\thesectionc.\thesubsectionc.\thesubsubsectionc.
    {\kern1pt \tenit #1}}\par\vspace{5pt}}
\newcommand{\nonumsection}[1] {\vspace{12pt}\noindent{\tenbf #1}
    \par\vspace{5pt}}

\newcounter{appendixc}
\newcounter{subappendixc}[appendixc]
\newcounter{subsubappendixc}[subappendixc]
\renewcommand{\thesubappendixc}{\Alph{appendixc}.\arabic{subappendixc}}
\renewcommand{\thesubsubappendixc}
    {\Alph{appendixc}.\arabic{subappendixc}.\arabic{subsubappendixc}}

\renewcommand{\appendix}[1] {\vspace{12pt}
        \refstepcounter{appendixc}
        \setcounter{figure}{0}
        \setcounter{table}{0}
        \setcounter{lemma}{0}
        \setcounter{theorem}{0}
        \setcounter{corollary}{0}
        \setcounter{definition}{0}
        \setcounter{equation}{0}
        \renewcommand{\thefigure}{\Alph{appendixc}.\arabic{figure}}
        \renewcommand{\thetable}{\Alph{appendixc}.\arabic{table}}
        \renewcommand{\theappendixc}{\Alph{appendixc}}
        \renewcommand{\thelemma}{\Alph{appendixc}.\arabic{lemma}}
        \renewcommand{\thetheorem}{\Alph{appendixc}.\arabic{theorem}}
        \renewcommand{\thedefinition}{\Alph{appendixc}.\arabic{definition}}
        \renewcommand{\thecorollary}{\Alph{appendixc}.\arabic{corollary}}
        \renewcommand{\theequation}{\Alph{appendixc}.\arabic{equation}}
        \noindent{\tenbf Appendix \theappendixc #1}\par\vspace{5pt}}
\newcommand{\subappendix}[1] {\vspace{12pt}
        \refstepcounter{subappendixc}
        \noindent{\bf Appendix \thesubappendixc. {\kern1pt \bfit #1}}
    \par\vspace{5pt}}
\newcommand{\subsubappendix}[1] {\vspace{12pt}
        \refstepcounter{subsubappendixc}
        \noindent{\rm Appendix \thesubsubappendixc. {\kern1pt \tenit #1}}
    \par\vspace{5pt}}

\newcommand{\textlineskip}{\baselineskip=13pt}
\newcommand{\smalllineskip}{\baselineskip=10pt}

\def\eightcirc{
\begin{picture}(0,0)
\put(4.4,1.8){\circle{6.5}}
\end{picture}}
\def\eightcopyright{\eightcirc\kern2.7pt\hbox{\eightrm c}}

\newcommand{\copyrightheading}[1]
    {\vspace*{-2.5cm}\smalllineskip{\flushleft
    {\footnotesize International Journal of Modern Physics A #1}\\
    {\footnotesize $\eightcopyright$\, World Scientific Publishing
     Company}\\
     }}

\def\abstracts#1#2#3{{
    \centering{\begin{minipage}{4.5in}\footnotesize\baselineskip=10pt
    \parindent=0pt #1\par
    \parindent=15pt #2\par
    \parindent=15pt #3
    \end{minipage}}\par}}

\renewenvironment{thebibliography}[1]
    {\frenchspacing
     \ninerm\baselineskip=11pt
     \begin{list}{\arabic{enumi}.}
    {\usecounter{enumi}\setlength{\parsep}{0pt}
     \setlength{\leftmargin 12.7pt}{\rightmargin 0pt} 
     \setlength{\itemsep}{0pt} \settowidth
    {\labelwidth}{#1.}\sloppy}}{\end{list}}

\newcounter{itemlistc}
\newcounter{romanlistc}
\newcounter{alphlistc}
\newcounter{arabiclistc}

\newcommand{\fcaption}[1]{
        \refstepcounter{figure}
        \setbox\@tempboxa = \hbox{\footnotesize Fig.~\thefigure. #1}
        \ifdim \wd\@tempboxa > 5in
           {\begin{center}
        \parbox{5in}{\footnotesize\smalllineskip Fig.~\thefigure. #1}
            \end{center}}
        \else
             {\begin{center}
             {\footnotesize Fig.~\thefigure. #1}
              \end{center}}
        \fi}

\newcommand{\tcaption}[1]{
        \refstepcounter{table}
        \setbox\@tempboxa = \hbox{\footnotesize Table~\thetable. #1}
        \ifdim \wd\@tempboxa > 5in
           {\begin{center}
        \parbox{5in}{\footnotesize\smalllineskip Table~\thetable. #1}
            \end{center}}
        \else
             {\begin{center}
             {\footnotesize Table~\thetable. #1}
              \end{center}}
        \fi}

\def\@citex[#1]#2{\if@filesw\immediate\write\@auxout
    {\string\citation{#2}}\fi
\def\@citea{}\@cite{\@for\@citeb:=#2\do
    {\@citea\def\@citea{,}\@ifundefined
    {b@\@citeb}{{\bf ?}\@warning
    {Citation `\@citeb' on page \thepage \space undefined}}
    {\csname b@\@citeb\endcsname}}}{#1}}

\newif\if@cghi
\def\cite{\@cghitrue\@ifnextchar [{\@tempswatrue
    \@citex}{\@tempswafalse\@citex[]}}
\def\citelow{\@cghifalse\@ifnextchar [{\@tempswatrue
    \@citex}{\@tempswafalse\@citex[]}}
\def\@cite#1#2{{$\null^{#1}$\if@tempswa\typeout
    {IJCGA warning: optional citation argument
    ignored: `#2'} \fi}}

\def\pmb#1{\setbox0=\hbox{#1}
    \kern-.025em\copy0\kern-\wd0
    \kern.05em\copy0\kern-\wd0
    \kern-.025em\raise.0433em\box0}

\def\fnt#1#2{\footnotetext{\kern-.3em
    {$^{\mbox{\scriptsize #1}}$}{#2}}}

\def\@makefnmark{\hbox to 0pt{$^{\@thefnmark}$\hss}}    

\def\ps@myheadings{%
    \let\@oddfoot\@empty\let\@evenfoot\@empty
    \def\@evenhead{\slshape\leftmark\hfil}
    \def\@oddhead{\hfil{\slshape\rightmark}}
    \let\@mkboth\@gobbletwo
    \let\sectionmark\@gobble
    \let\subsectionmark\@gobble
    }
\font\tenrm=cmr10
\font\tenit=cmti10
\font\tenbf=cmbx10
\font\bfit=cmbxti10 at 10pt
\font\ninerm=cmr9

\font\eightrm=cmr8

\def\qed{\hbox{${\vcenter{\vbox{            
   \hrule height 0.4pt\hbox{\vrule width 0.4pt height 6pt
   \kern5pt\vrule width 0.4pt}\hrule height 0.4pt}}}$}}

\def\slashchar#1{\setbox0=\hbox{$#1$}           
   \dimen0=\wd0                                 
   \setbox1=\hbox{/} \dimen1=\wd1               
   \ifdim\dimen0>\dimen1                        
      \rlap{\hbox to \dimen0{\hfil/\hfil}}      
      #1                                        
   \else                                        
      \rlap{\hbox to \dimen1{\hfil$#1$\hfil}}   
      /                                         
   \fi}                                         %
\newcommand{\blank}{}
\renewcommand{\theequation}{\blank \arabic{equation}}
\newcounter{dummy}{}
\newcommand{\letters}{%
    \setcounter{dummy}{\value{equation}}
    \renewcommand{\thedummy}{\blank \arabic{dummy}}
    \renewcommand{\theequation}{\thedummy\alph{equation}}
    \refstepcounter{dummy}
    \setcounter{equation}{0}%
 }
\newcommand{\noletters}{%
    \setcounter{equation}{\value{dummy}}
    \renewcommand{\theequation}{\blank\arabic{equation}}%
  }
\newenvironment{mathletters}{\letters}{\noletters}
\newcommand{\half}{\frac 12}
\newcommand{\sgn}{\mathop{\mathrm{sgn}}}
\newcommand{\diag}{\mathop{\mathrm{diag}}}

\begin{document}
\setlength{\textheight}{7.7truein}  
\thispagestyle{empty}
\markboth{\protect{\footnotesize\it A.~A.~Abrikosov, jr.}}%
{\protect{\footnotesize\it Fermion States on the Sphere $S^2$}}
\normalsize\textlineskip
\setcounter{page}{1}
\copyrightheading{}     
\vspace*{0.88truein}

\centerline{\bf FERMION STATES ON THE SPHERE $S^2$}
\vspace*{0.37truein}
\centerline{\footnotesize A.~A.~ABRIKOSOV, JR.\footnote{PERSIK@VXITEP.ITEP.RU.}}
\baselineskip=12pt
\centerline{\footnotesize\it Institute of Theoretical and Experimental Physics,}
\baselineskip=10pt
\centerline{\footnotesize\it B.~Cheremushkinskaya~Str.,~25,}
\centerline{\footnotesize\it Moscow, 117218, Russia}
\vspace*{0.225truein}
\abstracts{We solve for the spectrum and eigenfunctions of Dirac operator on
the sphere. The eigenvalues are nonzero whole numbers. The eigenfunctions
are two-component spinors which may be classified by representations of the
$SU(2)$ group with half-integer angular momenta. They are special linear
combinations of conventional spherical spinors. }%
{\sc keywords:}{\it Dirac operator,  spherical spinors, spherical functions,
Jacobi polynomials}
%
\vspace*{1pt}\textlineskip  
\nonumsection{Introduction}   
\vspace*{-0.5pt}
The reasons that inspired us to address the problem were primarily physical.
There are at least two fields where the results can be applied. The first is
the bag model with spectral boundary conditions\cite{dowker/etal}. This may
be of interest for physics of strong interactions. On the other hand it is
known that electrons in extensively studied currently fulleren molecules
(such as $\mathrm{C}_{60}$ and others) obey the Dirac equation. Our problem
bears the direct relation to the continuous limit of electronic
states in fullerens.

Surprisingly but despite all its beauty the problem escaped textbooks. A
general construction of the eigenfunctions of Dirac operator on
$N$-dimensional spheres was given in paper\cite{camporesi/higuchi}. It was
shown that those can be written in terms of Jacobi polynomials. Our present
goal was to study in detail and find explicit formulas for the particular
case and to emphasize the group properties of solutions. We shall start from
approaching the eigenvalue problem in Sect.~2 and construct the
$SU(2)$-algebra in Sect.~3. This makes possible to classify the obtained in
Sect.~4 solutions under the group representations. Finally after making a
bridge to conventional spherical spinors in Sect.~5 we shall summarize the
results at the end.

\section{The Dirac operator}

First of all we shall introduce the notation and  then write down the Dirac
operator. We shall use the standard parameterization of the unit sphere
$S^2$:
\begin{equation} \label{sph-coord}
 x = \sin \theta \cos \phi ;   \quad
 y = \sin \theta \sin \phi ;   \quad
 z = \cos \theta ;
\end{equation}
Spinors in two dimensions have two components and the role of Dirac matrices
belongs to Pauli matrices: $(\gamma^1,\, \gamma^2) \rightarrow (\sigma_x,\,
\sigma_y)$. Dirac operator is a convolution of covariant derivatives in
spinor representation with zweibein
 $e^{\alpha \,a}= \diag (1,\, \sin^{-1} \theta)$
and $\sigma$-matrices:
\begin{equation}\label{Dirac}
  -i \hat{\nabla} = -i\, e^{\alpha\,a} \sigma_a \nabla_\alpha =
    -i \sigma_x
    \left(\partial_\theta + \frac{\cot\theta}2 \right)
    -\frac{i \sigma_y}{\sin\theta}\, \partial_\phi .
\end{equation}

The general theorem (which also applies to the sphere) states that Dirac
operator has no zero eigenvalues on manifolds with positive curvature.
Therefore there must be no zero value in the spectrum of operator
(\ref{Dirac}).

\section{The eigenvalue problem} \label{eigenvalue}

Eigenfunctions of Dirac operator are two-component spinors
  $\psi_\lambda(\theta,\, \phi)$ that satisfy the equation:
\begin{equation}\label{DiracS2}
 -i\,\hat{\nabla}\psi_\lambda(\theta,\, \phi) =
    \lambda\, \psi_\lambda(\theta,\, \phi).
\end{equation}
Expanding them into Fourier series
  $\psi_\lambda(\theta,\, \phi) =
    (2\pi)^{-\half} \sum_m \psi_{\lambda m}(\theta) \exp i\, m\, \phi $
we obtain independent equations for the components.  Obviously because of
$\psi_\lambda$ being spinors the sum runs over all half-integer values of
$m$.

The square of $-i \hat{\nabla}$ is a diagonal operator.  After the change of
variables  $x =\cos \theta$, $x \in [-1,\, 1]$ we obtain separate
generalized hypergeometric equations for the upper and lower components
  $\alpha_{\lambda m}(x)$ and $\beta_{\lambda m}(x)$:
\begin{equation}\label{ODEf}
  \left[\frac d{dx}\, (1-x^2)\, \frac d{dx} -
    \frac{m^2 - \sigma_z\, m\, x +\frac 14}{1-x^2}\right] \left(
    \begin{array}{c}
      \alpha_{\lambda m} \\
      \beta_{\lambda m}
    \end{array}\right) =
    - \left(\lambda^2 -\frac 14\right)\, \left(
    \begin{array}{c}
      \alpha_{\lambda m} \\
      \beta_{\lambda m}
    \end{array}\right).
\end{equation}
Because of $\sigma_z = \diag (1,\, -1)$ in the second term the equations for
$\alpha$ and $\beta$ differ. Inversion $x \rightarrow -x$ (or, equivalently,
$m \rightarrow -m$) transforms one into another.

A regular routine brings (\ref{ODEf}) to equations of hypergeometric type.
Those have square integrable on the interval $x \in [-1,\, 1]$ solutions
provided that
\begin{equation}\label{lmb=m+n+}
  \lambda^2 = \left(n + |m| + \half\right)^2,
\end{equation}
with integer $n \geq 0$. Thus $\lambda = \pm 1,\, \pm 2 \dots$ are nonzero
integers and indeed there are no zero-modes of Dirac operator on the sphere.
The solutions may be expressed in terms of Jacobi polynomials of $n$-th
order, like in \cite{camporesi/higuchi}.  We shall put them into another
form that will be given later, see Eqs.~(\ref{ups>}).

\section{The $SU(2)$ algebra} \label{SU(2)}
Let us show that Dirac operator on the sphere $S^2$ is invariant under
transformations of the $SU(2)$ group. Its algebra consists of three
operators:
\begin{mathletters}\label{L-ops}
\begin{eqnarray}
  \hat{L}_z & = &
    - i \frac \partial{\partial \phi}; \label{L-ops/a}\\
  \hat{L}_+ & = &
    \hphantom{-} e^{\hphantom{-} i \phi}
    \left(\frac \partial{\partial \theta}
    + i\, \cot \theta \frac \partial{\partial \phi}
    + \frac{\sigma_z}{2\, \sin \theta} \right); \label{L-ops/b}\\
  \hat{L}_- & = &
   - e^{- i \phi}
    \left(\frac \partial{\partial \theta}
    - i\, \cot \theta \frac \partial{\partial \phi}
    - \frac{\sigma_z}{2\, \sin \theta} \right).  \label{L-ops/c}
\end{eqnarray}
\end{mathletters}
The operators satisfy the standard commutation relations of $SU(2)$
algebra:
\begin{equation} \label{L-comm}
  \left[ \hat{L}_z ,\, \hat{L}_+ \right] =   \hat{L}_+ ;
\qquad
  \left[ \hat{L}_z ,\, \hat{L}_- \right] = - \hat{L}_- ;
\qquad
  \left[ \hat{L}_+ ,\, \hat{L}_- \right] = 2 \hat{L}_z .
\end{equation}

The direct check proves that they also commute with Dirac operator (\ref{Dirac}).
Hence its eigenfunctions may be classified by representations of the
$SU(2)$-group. Action of the generators $\hat{L}_+$ ($\hat{L}_-$) raises
(lowers) the value of $m$ leaving $\lambda$ intact.

The spherical d'Alembert operator is directly linked to the square of
angular momentum:
\begin{equation}\label{nabla/L}
  - \hat{\nabla}^2 = \hat{L}^2 + \frac 14
    =\hat{L}_z^2 +
    \half(\hat{L}_+ \hat{L}_- + \hat{L}_- \hat{L}_+) +\frac 14.
\end{equation}
This results into the following relation between the eigenvalues:
\begin{equation}\label{L-sqr:mn}
  \langle m,\, n | \hat{L}^2 | m,\, n \rangle = \lambda^2 - \frac 14
    = (n + |m|)(n + |m| + 1).
\end{equation}
The proper values of angular momentum are half-integers
  $l = |\lambda| - \half = n + |m|$.

Operators (\ref{L-ops}) are diagonal. Hence formally $\alpha$ and $\beta$
belong to different representations. Nevertheless they form doublets with
respect to the Dirac operator.

\section{The spinor spherical functions} \label{Upsilon}

We shall list spinor spherical functions $\Upsilon^{\varepsilon}_{l m}$ by
the values of angular momentum  $l = n + |m|$, its $z$-projection $m$ and
  $\varepsilon = \sgn \lambda$.
Let us introduce the integers
  $l^\pm = l \pm \half$ and $m^\pm = m \pm \half$.
Using the $\pm$-superscripts for $\varepsilon$ we may write:
\begin{eqnarray} \label{ups>}
  \lefteqn{\Upsilon^\pm_{lm} (x,\, \phi) =
    \pm \frac{i^{l^+}\, (-1)^{l^-}}{2^{l^+} \Gamma(l^+)}
    \sqrt{\frac{(l + m)!}{(l - m)!}}} & &
    \nonumber \\
    & \times &
    \frac{e^{i m \phi}}{\sqrt{2\pi}}
    \left(
    \begin{array}{r}
     \sqrt{\mp i} (1-x)^{-\frac{m^-}2} (1+x)^{-\frac{m^+}2}\,
    \frac{d^{l-m}}{dx^{l-m}} (1-x)^{l^-} (1+x)^{l^+}  \\
      \sqrt{\pm i} (1-x)^{-\frac{m^+}2} (1+x)^{-\frac{m^-}2}\,
    \frac{d^{l-m}}{dx^{l-m}} (1-x)^{l^+} (1+x)^{l^-}
    \end{array}\right).
  \end{eqnarray}
This representation of $SU(2)$ multiplets was constructed by successive
application of operator $\hat{L}_-$ to the function with maximum projection
$m = l$ of momentum. The constants were chosen in order to ensure the
correct behaviour of spinors under complex conjugation\cite{Landau/Lifshitz}
and fix the right signs of matrix elements,
\begin{mathletters}\label{UpsLUps}
\begin{eqnarray}
   \langle l,\, m-1\,  |\hat{L}_-|\,  l,\, m \rangle & = &
    \langle l,\, m \, |\hat{L}_+|\,  l,\, m-1 \rangle =
    \sqrt{(l+m)(l-m+1)}
     \label{UpsLUps/a}; \\
    \langle l,\, m \, |\hat{L}_z|\,  l,\, m \rangle &  =  & m.
    \label{UpsLUps/b}
\end{eqnarray}
\end{mathletters}
The spinor spherical functions constitute a complete orthonormal system:
\begin{equation}\label{onorm}
  \langle \Upsilon^{\varepsilon_1}_{l_1 m_1} |
    \Upsilon^{\varepsilon_2}_{l_2 m_2}\rangle =
    \int_0^{2\pi} d\phi \int_0^\pi
    (\Upsilon^{\varepsilon_1}_{l_1 m_1})^\dagger
    \Upsilon^{\varepsilon_2}_{l_2 m_2}\,
    \sin \theta\, d\theta =
    \delta^{\varepsilon_1 \varepsilon_2}\,
    \delta_{l_1 l_2}\, \delta_{m_1 m_2}.
\end{equation}
 This makes them quite a useful means of harmonic analysis on $S^2$.

\section{Relation to the ordinary spherical spinors} \label{Ups<->Omega}

Conventional spherical spinors $\Omega_{j,\, l,\, m}$ are characterized by
total angular momentum $j$, orbital angular momentum $l$ and $z$-projection
of the total momentum $m$. They also form on $S^2$ a complete functional
system that differs from $\Upsilon^\varepsilon_{l m}$. The reason is that
$\Omega$-spinors were constructed in the flat $3d$-space using the set of
$\sigma$-matrices aligned with Cartesian frame whereas our
$\gamma$-matrices were specific to the curved $2d$-manifold $S^2$.
Certainly the two types of spinors must be interrelated.

Transformation of spinors from spherical to Cartesian coordinates includes
multiplication by matrix
   $V^\dagger = \exp -\frac{i\sigma_z}2 \phi
    \exp -\frac{i\sigma_y}2 \theta$.
This leads to the relation:
\begin{equation}\label{V-Ups}
  V^\dagger\, \Upsilon^\pm_{l\, m} =
    \frac 1{\sqrt 2}
    \left(
\begin{array}{c}
    \sqrt{\frac{l+m}{2l}} \, Y_{l^- m^-} \pm
    \sqrt{\frac{l-m+1}{2l+2}} \, Y_{l^+ m^-}
     \\
    \sqrt{\frac{l-m}{2l}} \, Y_{l^- m^+} \mp
    \sqrt{\frac{l+m+1}{2l+2}} \, Y_{l^+ m^+}
\end{array}
    \right) =
    \frac 1{\sqrt 2}
    \left( \Omega_{l,\, l^-,\, m} \mp \Omega_{l,\, l^+,\, m} \right).
\end{equation}
Thus $\Upsilon$'s are combined of $\Omega$'s with the same values of $j$ and
$m$. Therefore our functions have definite values of total angular
momentum $\hat{J}$ and its $z$-projection $\hat{J}_z$:
\begin{equation}\label{j,m:Ups}
  \hat{\mathbf{J}}^2 \, \Upsilon_{lm} =
    (\hat{\mathbf{L}}_C+\hat{\mathbf{S}}_C)^2 \, \Upsilon_{lm} =
    l (l+1) \, \Upsilon_{lm}
    \qquad {\mathrm{and}} \qquad
    \hat{J}_z \, \Upsilon_{lm} = m  \, \Upsilon_{lm};
\end{equation}
where $\hat{\mathbf{L}}_C$ and $\hat{\mathbf{S}}_C$ are the Cartesian
operators of orbital momentum and spin vectors respectively. In the mean
time $\Upsilon^\pm_{lm}$ do not diagonalize orbital momentum $\hat l$.

\section{Summary}

We have shown that the spectrum of Dirac operator on the sphere consists of
nonzero integers. The eigenfunctions are two-component spherical spinors
that can be grouped in $SU(2)$ multiplets with half-integer values of total
momentum. Our solutions differ from customary spherical spinors obtained in
the flat space being the linear combinations of those.

\nonumsection{Acknowledgements}
 \noindent
But to acknowledging the traditionally excellent work of the organizers I
would like to express my deep gratitude to the ``Fundamental Nuclear
Physics'' program of Russian Ministry of Science and the Organizing
Committee for financial support of my participation in the Workshop. The
work was done with support from the RFBR grant 00--02--17808.

\nonumsection{References}

\end{document}